\documentclass[twocolumn,prb,aps,superscriptaddress,showpacs,amsmath,amssymb]{revtex4}
\usepackage{amsfonts}
\usepackage{bbm}
\usepackage{tipa}

\usepackage{graphicx}% Include figure files
\usepackage{dcolumn}% Align table columns on decimal point
\usepackage{bm}% bold math
\begin{document}
%\preprint{}

\title{Effect of gate voltage on spin transport along $\alpha$-helical protein}
\author{Ting-Rui Pan}
\affiliation{International Center for Quantum Materials, School of Physics, Peking University, Beijing 100871, China}
\author{Ai-Min Guo}
\affiliation{Department of Physics, Harbin Institute of Technology, Harbin 150001, China}
\author{Qing-Feng Sun}
\email{sunqf@pku.edu.cn}
\affiliation{International Center for Quantum Materials, School of Physics, Peking University, Beijing 100871, China}
\affiliation{Collaborative Innovation Center of Quantum Matter, Beijing 100871, China}
\date{\today}

\begin{abstract}
Recently, the chiral-induced spin selectivity in molecular systems has attracted extensive interest among the scientific communities. Here, we investigate the effect of the gate voltage on spin-selective electron transport through the $\alpha$-helical peptide/protein molecule contacted by two nonmagnetic electrodes. Based on an effective model Hamiltonian and the Landauer-B\"{u}ttiker formula, we calculate the conductance and the spin polarization under an external electric field which is perpendicular to the helix axis of the $\alpha$-helical peptide/protein molecule. Our results indicate that both the magnitude and the direction of the gate field have a significant effect on the conductance and the spin polarization. The spin filtration efficiency can be improved by properly tuning the gate voltage, especially in the case of strong dephasing regime. And the spin polarization increases monotonically with the molecular length without the gate voltage, which is consistent with the recent experiment, and presents oscillating behavior in the presence of the gate voltage. In addition, the spin selectivity is robust against the dephasing, the on-site energy disorder, and the space angle disorder under the gate voltage. Our results could motivate further experimental and theoretical works on the chiral-based spin selectivity in molecular systems.
\end{abstract}


%Use showkeys class option if keyword
%display desired
\maketitle

\section{\label{sec1}Introduction}

Molecular spintronics, which manipulates the electron spin transport through organic molecules, has been receiving more and more attentions, because of the potential applications in storage and processing of information.\cite{fa, rar, ss, dva, nwjm, bj, wfj} A growing number of pioneering researches have established organic materials as a viable platform for spin-filter devices. The experimental evidence of room-temperature spin-polarized injection and transport through prototypical organic semiconductors was demonstrated for the first time.\cite{dv} An original spin-valve device was designed in which a single-walled carbon nanotube is laterally coupled to single-molecule magnets through supramolecular interactions.\cite{um} Spin-selective carrier transmission was demonstrated in a nonmagnetic system which is composed of carbon nanotube-DNA hybrid.\cite{dgs, akm} Both self-assembled monolayers of double-stranded DNA (dsDNA) deposited on gold substrate\cite{gb} and single dsDNA molecules connected by two electrodes\cite{xz} presented high spin polarization at room temperature. However, spin-polarized effect was not observed in single-stranded DNA monolayers.\cite{gb} Later on, spin-selective effect of electron transmission along bacteriorhodopsin---an $\alpha$-helical protein---embedded in purple membrane physisorbed on gold and aluminum substrates was demonstrated.\cite{md} Some works theoretically investigated the spin transport properties of the chiral molecular systems.\cite{gr,me,eaa,rd,ys,kks} A model Hamiltonian, including the small environment-induced dephasing, the weak spin-orbit coupling (SOC), and the helical symmetry, was proposed to perfectly rationalize the quantum spin transport through the dsDNA and the single-stranded DNA,\cite{gam1,gam5} as well as the $\alpha$-helical protein, and meanwhile explained the contradictory results between the protein and the single-stranded DNA\cite{gam2}. On the basis of self-assembled monolayer of $\alpha$-helical polyalanine adsorbed on gold, a device was presented and indicated the ability to produce spin-based device without a permanent magnet.\cite{dob} In addition, it was demonstrated that the electrons transmitted through Photosystem I which is mainly composed of protein complex were highly spin-selective and the spin polarization was temperature-dependent.\cite{ci} Very recently, the light-controlled ability on spin filtration through the bacteriorhodopsin D96N mutant was observed.\cite{eh} The spin selectivity of electron was measured through the monolayers of oligopeptides and  increased with the increase of the molecular length in the investigated length range.\cite{km}

Recent experimental and theoretical researches on spin-dependent electron transport through the DNA and protein molecules have given rise to a prominent improvement of molecular spintronics. Electron transport through the dsDNA molecules presented high spin polarization. Meanwhile, the dsDNA molecules could act as a field-effect transistor under a gate voltage.\cite{ykh} Then a theoretical investigation on the influence of gate voltage on spin transmission along the dsDNA molecules was performed, revealing that spin polarization showed strong dependence on the magnitude as well as the direction of gate voltage and could be significantly enhanced by tuning the gate voltage.\cite{gam3} As for the $\alpha$-helical protein, the transmitted electrons also exhibited the ability of spin filtration. In the presence of gate voltage, protein field-effect transistors were also reported and a model for transport was proposed.\cite{mg,bk} Based on the similarly unique helical structure, spin-selective properties of the $\alpha$-helical protein were compared with that of the DNA molecules and then a question came to us that whether the gate voltage has also such a intense effect on the spin transport along $\alpha$-helical protein, just like along DNA molecules. In addition, we also wanted to investigate how the on-site energy disorder and space angle disorder affect spin polarization of the peptide in the presence of gate voltage.

Herein, we report on a method to regulate the spin-dependent electron transport along the $\alpha$-helical protein molecule connected by two non-magnetic electrodes in the presence of gate voltage, which gives rise to an external electric field perpendicular to the helix axis of the molecule, as illustrated in Fig.~\ref{fig:zero}. On the basis of an effective model Hamiltonian and the Landauer-B\"{u}ttiker formula, the conductance and the spin polarization are calculated. Our results indicate that the spin filtration efficiency shows strong dependence on the magnitude as well as the direction of the gate voltage. One can improve the spin polarization by properly tuning the gate voltage, especially in the case of strong dephasing regime, such as the high experimental temperature. Both the conductance and the spin polarization versus the protein length show oscillating behavior when the gate voltage is employed. We also find that the spin polarization of the peptide is robust against the dephasing, the on-site energy disorder and the space angle disorder under the gate voltage. Therefore the $\alpha$-helical protein-based device could be a more efficient spin filter by properly tuning the gate voltage.

The rest of the paper is organized as follows. In Sec.~\ref{sec2}, the calculation model and the method are presented. In Sec.~\ref{sec3}, the effect of the gate voltage on the conductance and the spin polarization is shown. And then, we investigate the influence of the on-site energy disorder and the space angle disorder. Finally, the results are summarized in Sec.~\ref{sec4}.

\section{\label{sec2}Model and method}

The spin transport properties of the $\alpha$-helical protein can be simulated by the Hamiltonian:\cite{gam2}
\begin{eqnarray}
\begin{aligned}
{\cal H}= & \left(\sum_{n=1}^N \varepsilon_ {n} c_{n}^\dag c_{n}+ \sum _{n=1}^{N-1}\sum _{j=1}^{N-n} t_{nj}c_{n}^\dag c_{n+j}  +\mathrm{H.c.}\right) \\& + \sum_{n=1}^{N-1}\sum_{j=1}^{N-n} ( 2i \mu_{nj} c_{n}^\dag \sigma _{nj}c_{n+j}+ \mathrm{H.c.}) \\& +\sum_{n<1} ( \varepsilon_ {m} b_{n}^\dag b_{n}+t_ {m} b_{n}^ \dag b_{n-1}+\mathrm{H.c.})\\&+\sum_{n>N} ( \varepsilon_ {m} b_{n}^\dag b_{n}+t_ {m}b_{n}^\dag b_{n+1} +\mathrm{H.c.}) \\ & + \tau( b_{0} ^ \dag c_{1} + c_{N} ^ \dag b_{N+1}+ \mathrm {H.c.})  \\ & + \sum_{n=1}^{N}\sum_{k} (\varepsilon_{n k} a_{n k}^\dag a_{n k}+t_d a_{n k}^\dag c_{n}+ \mathrm {H.c.}). \label{eq:one}
\end{aligned}
\end{eqnarray}
The first two terms are the Hamiltonian of the $\alpha$-helical protein whose length is $N$, with $c_{n} ^\dag= (c_{n \uparrow} ^\dag, c_{ n \downarrow } ^\dag)$ and $c_{n} = (c_{n \uparrow} , c_{ n \downarrow } )^{\rm T}$ being the creation and annihilation operators, respectively. $\varepsilon_ {n}$ is the on-site energy, $t_{nj}=t_1e^{-(l_{nj}-l_1)/l_c}$ is the hopping integral between two neighboring sites $n$ and $n+j$, $\mu_{nj}=s_1\cos (\varphi_{nj}^-) e^{- (l_{nj} -l_1)/l_c}$ is the corresponding SOC, and $\sigma _{nj}=(\sigma _{x}\sin {\varphi_{nj}^+-\sigma _{y}} \cos {\varphi_{nj}^+})\sin {\theta_{nj}}+ \sigma_z\cos{\theta_{nj}}$.\cite{gam2} Here, $l_{nj}=\sqrt{[2R \sin(\varphi_{nj}^-)] ^2+(h_{n+j} -h_{n})^2}$ is the Euclidean distance between sites $n$ and $n+j$, $\varphi_{nj} ^{\pm} = (\varphi_{n+j} \pm\varphi_ {n})/2$, and $\theta_{nj} =\arccos [2R\sin(\varphi_{nj}^-)/l_{nj}]$ is the space angle, where the parameters ($R$, $\varphi_n$, $h_n$) are the three cylindrical coordinates of site $n$. $\sigma_{x,y,z}$ are the Pauli matrices, $l_c$ is the decay exponent, and $s_1$ is the renormalized SOC. In the absence of any disorder, $l_{nj}$ for $j=1$ is reduced to $l_1=\sqrt{[2R\sin (\Delta \varphi/2)]^2+(\Delta h)^2}$ and correspondingly $t_{nj}$ for $j=1$ is reduced to $t_1$ which denotes the hopping integral between the nearest neighbor (NN) sites, with $\Delta \varphi$ and $\Delta h$ being the twist angle and the stacking distance between the NN sites, respectively. The third and the fourth terms are the Hamiltonians of the left and right semi-infinite real electrodes, respectively. The next one represents the couplings between the $\alpha$-helical protein and the two real electrodes. Finally, the last term denotes the B\"{u}ttiker's virtual electrodes and their couplings to each site of the molecule, which is introduced to describe the dephasing processes caused by the electrons' inelastic scatterings with the electrons, the phonons, the counterions, and the adsorbed impurities.\cite{xy,jh} Actually, previous works have clearly indicated the presence of the phase-breaking processes in the proteins.\cite{sss, gb1, mt}

\begin{figure}
\includegraphics[scale=0.31]{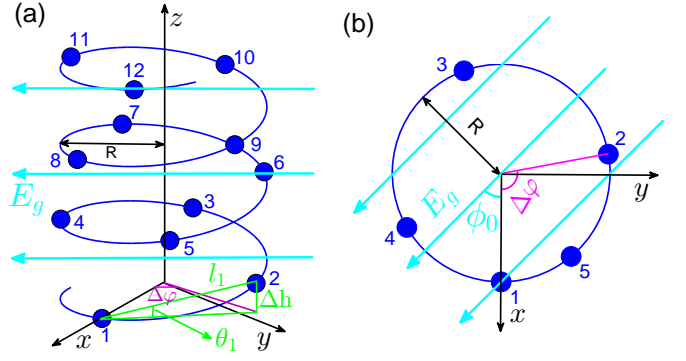}
\caption{\label{fig:zero}(color online). (a) Schematic view of the $\alpha$-helical protein under external electric field $E_g$ which is perpendicular to the helix axis ($z$-axis). The parameters are radius $R=0.25$ nm,  Euclidean distance $l_1=0.41$ nm, space angle $\theta_1=0.37$ rad, twist angle $\Delta\varphi=5\pi/9$, and stacking distance $\Delta h=0.15$ nm between two nearest neighbour amino acids. The amino acids are labelled by Arabic numerals from the bottom up. (b) Projection of the five bottom amino acids and the electric field into the $x$-$y$ plane. Here, $\phi_0$ is the angle between the direction of the electric field and the positive $x$ axis direction. }
\end{figure}

When the $\alpha$-helical protein is subjected to an external electric field which is perpendicular to its helix axis (see Fig.~\ref{fig:zero}), the on-site energy at the $n$th site will be changed into the following form:
\begin{equation}
\varepsilon_ {n}=\varepsilon_ {n}^0+e E_g R \cos( \varphi_n + \phi_0), \label{eq:two}
\end{equation}
where $\varepsilon_ {n}^0$ is the on-site energy without the external electric field and $e$ is the elementary charge. $E_g$ is the perpendicular external electric field and thus the gate voltage across the $\alpha$-helical protein molecule is $2V_g=2E_gR$. The phase $\phi_0$, which is the angle from the direction of the external electric field to the positive direction of $x$-axis, shows the orientation of the gate voltage relative to the helical molecule, as seen in Fig.~\ref{fig:zero}. In order to adjust $\phi_0$, the helical molecule could be rotated with the direction of its helix axis fixed. Eq.~(\ref{eq:two}) shows that the gate voltage harmonically regulates the on-site energies along the helical strand and introduces periodic change of each site, due to the intrinsic helical structure of the protein. Undoubtedly, such adjustment will affect the electronic structure of the $\alpha$-helical molecule, which could make a significant effect on both the conductance and the spin polarization. The magnitude of the gate voltage chosen in this paper is the order of $0.1$ V, where the external electric field is much smaller than the internal one generated by the nuclei of the protein molecules and hence its effect on SOC may be negligible.

From the Landauer-B\"{u}ttiker formula, the current in the $q$th real or virtual electrode with spin $s= \uparrow,\downarrow$ can be described as $I_{q s}=(e^2/h) \sum_{m, s'}T_{qs,ms'} (V_{m}- V_{q})$, where $V_q$ is the voltage of the $q$th electrode and $T_{qs,ms'}$ is the transmission coefficient from the $m$th electrode with spin $s'$ to the $q$th electrode with spin $s$. With the boundary condition that the net current across each virtual electrode is zero, the voltage $V_q$ of the virtual electrodes can be derived by applying a small bias between the real electrodes with $V_L =V_b $ and $V_R=0$. Then the conductances of the right real electrode for spin-up ($G_{\uparrow}$) and spin-down ($G_{\downarrow}$) electrons can be written as $G_{s} =(e^2/h) \sum_{m, s'}T_{Rs,ms'} V_ {m}/ V_b $. Finally, the spin polarization is defined as:
\begin{equation}
P_s=\dfrac{G_ \uparrow- G_ \downarrow}{G_ \uparrow+ G_\downarrow}, \label{eq:three}
\end{equation}
and the averaged spin polarization is:
\begin{equation}
\quad\langle P_s \rangle=\dfrac{1}{\Omega}\int_{\Omega} P_s \, dE. \label{eq:three}
\end{equation}
Here, $\Omega$ denotes the lower energy band of $E<E_c$ and $E_c$ is the ``band center'', where the number of the electronic states below $E_c$ is equal to that above $E_c$.

For the $\alpha$-helical peptide, the structural parameters are the radius $R=0.25$ nm, the twist angle $\Delta\varphi=5\pi/9$, and the stacking distance $\Delta h=0.15$ nm. We take the NN hopping integral $t_1$ as energy unit. The values of aforementioned parameters are chosen as, the molecular length N=30, the on-site energy $\varepsilon_ {n}^0= 0$ without loss of universality, the renormalized SOC parameter $s_1=0.12t_1$, and the decay exponent $l_c=0.09$ nm.\cite{gam2}  For the real electrodes, the retarded self-energy can be numerically derived from $t_m=4t_1$ and $\tau=2t_1$.\cite{ldh} For the virtual electrodes, the dephasing strength is set to $\Gamma_d=0.02t_1$. The values of all above-mentioned parameters will be used throughout this paper except for specific annotation. We also investigate the spin transport through the $\alpha$-helical protein under the on-site energy disorder and the space angle disorder, as illustrated in Fig.~\ref{fig:seven}. The significant effect of the gate voltage on spin transport along the $\alpha$-helical protein molecule could be observed in a wide range of model parameters.

\section{\label{sec3}Results and Discussions}

\begin{figure}[b]
\center
\includegraphics[scale=0.37]{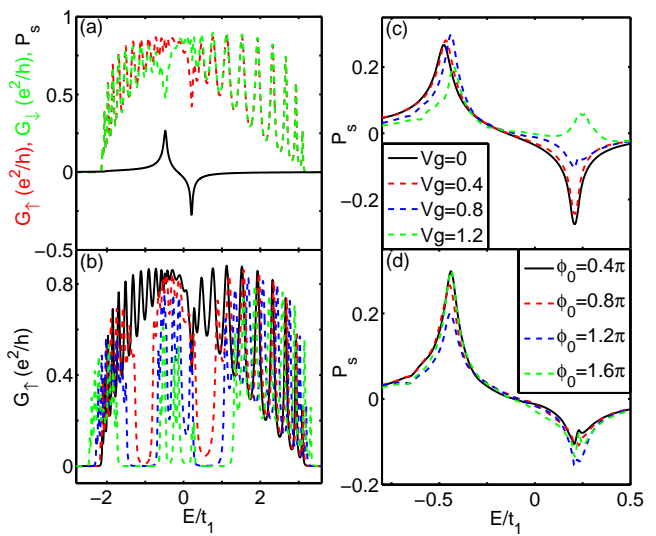}
\caption{\label{fig:one}(color online). (a) Energy-dependent spin-up conductance $G_{\uparrow}$ (red dashed line), spin-down one $G_{\downarrow}$ (green dashed line), and spin polarization $P_s$ (black solid line) for the $\alpha$-helical peptide without the gate voltage, i.e., $V_g=0$. (b) $G_{\uparrow}$ and (c) $P_s$ for the $\alpha$-helical peptide with different values of $V_g$ by fixing $\phi_0=0.4\pi$. The case of $V_g=0$ is also displayed as a comparison. The legend in (c) is for both (b) and (c).
(d) $P_s$ for several values of $\phi_0$ with $V_g=0.8$.}
\end{figure}

As a comparison, we first consider the spin transport through the $\alpha$-helical peptide with length $N=30$ in the absence of the gate voltage. Figure~\ref{fig:one}(a) shows the spin-up conductance $G_{\uparrow}$ (red-dashed line), the spin-down one $G_{\downarrow}$ (green-dashed line), and the spin polarization $P_s$ (black-solid line) with $V_g=0$ and $\Gamma_d=0.02t_1$. Although the dephasing strength is smaller than previous work,\cite{gam2} similar results can also be observed in the energy spectrum. For instance, there exist several sharp peaks in the curves of $G_{\uparrow}$-$E$ and $G_{\downarrow}$-$E$; the ``band center'' $E_c$ is shifted toward lower energy, i.e., $E_c<0$; except for the band center $E_c$ at which $P_s=0$, the spin polarization is nonzero and can achieve $\pm 26.7\%$, which is in accordance with the experimental result.\cite{md,km}

We then investigate the spin transport properties of the $\alpha$-helical peptide under the gate voltage. Figures~\ref{fig:one}(b) and \ref{fig:one}(c) display, respectively, the spin-up conductance $G_\uparrow$ and the spin polarization $P_s$ with several values of $V_g$ by fixing $\phi_0=0.4\pi$, while the other parameters are the same as Fig.~\ref{fig:one}(a). Here, the different colors of curves denote different $V_g$'s. In the presence of the gate voltage, the period of the on-site energies is changed to be eighteen amino acids, leading to the appearance of multiple potential barriers and wells within each period. As a result, the conductance is declined by increasing $V_g$, and several deep transmission valleys could emerge around $E_c$ and will develop into gaps in the case of larger $V_g$ [see the blue- and green-dashed lines in Fig.~\ref{fig:one}(b)]. Besides, the transmission spectrum can move toward both lower and higher energies by increasing $V_g$, since the range of the on-site energies $(-eV_g,eV_g)$ increases with $V_g$. Although the transmission ability of the peptide is weakened by $V_g$, its spin filtration efficiency is robust against $V_g$ and can even be enhanced by $V_g$ [Fig.~\ref{fig:one}(c)]. For instance, the maximum of $P_s$ is $26.7\%$, $28.1\%$, $29.9\%$, and $20.2\%$, respectively, by increasing $V_g$ from $0$, $0.4$, $0.8$, to $1.2$. Meanwhile, for $E<E_c$, the position of the peak becomes closer to $E_c$; for $E>E_c$, the position of the valley (peak) remains the same as that of $V_g=0$. This phenomenon is different from the dsDNA molecules.\cite{gam3} What's more, it is interesting that the sign of the spin polarization $P_s$ can be reversed in the case of large $V_g$ [see the green-dashed line in Fig.~\ref{fig:one}(c)]. Figure~\ref{fig:one}(d) shows $P_s$ for different values of $\phi_0$ by fixing $V_g=0.8$. Although the positions of the peak and the valley almost keep still by changing $\phi_0$, the spin filtration efficiency of the peptide considerably depends on $\phi_0$ and can be improved by properly modulating $\phi_0$. For instance, when $E=-0.44t_1$, $P_s$ is $29.9\%$, $26.1\%$, $20.0\%$, and $30.0\%$, respectively, by varying $\phi_0$ from $0.4\pi$, $0.8\pi$, $1.2\pi$, to $1.6\pi$.

\begin{figure}
\includegraphics[scale=0.47]{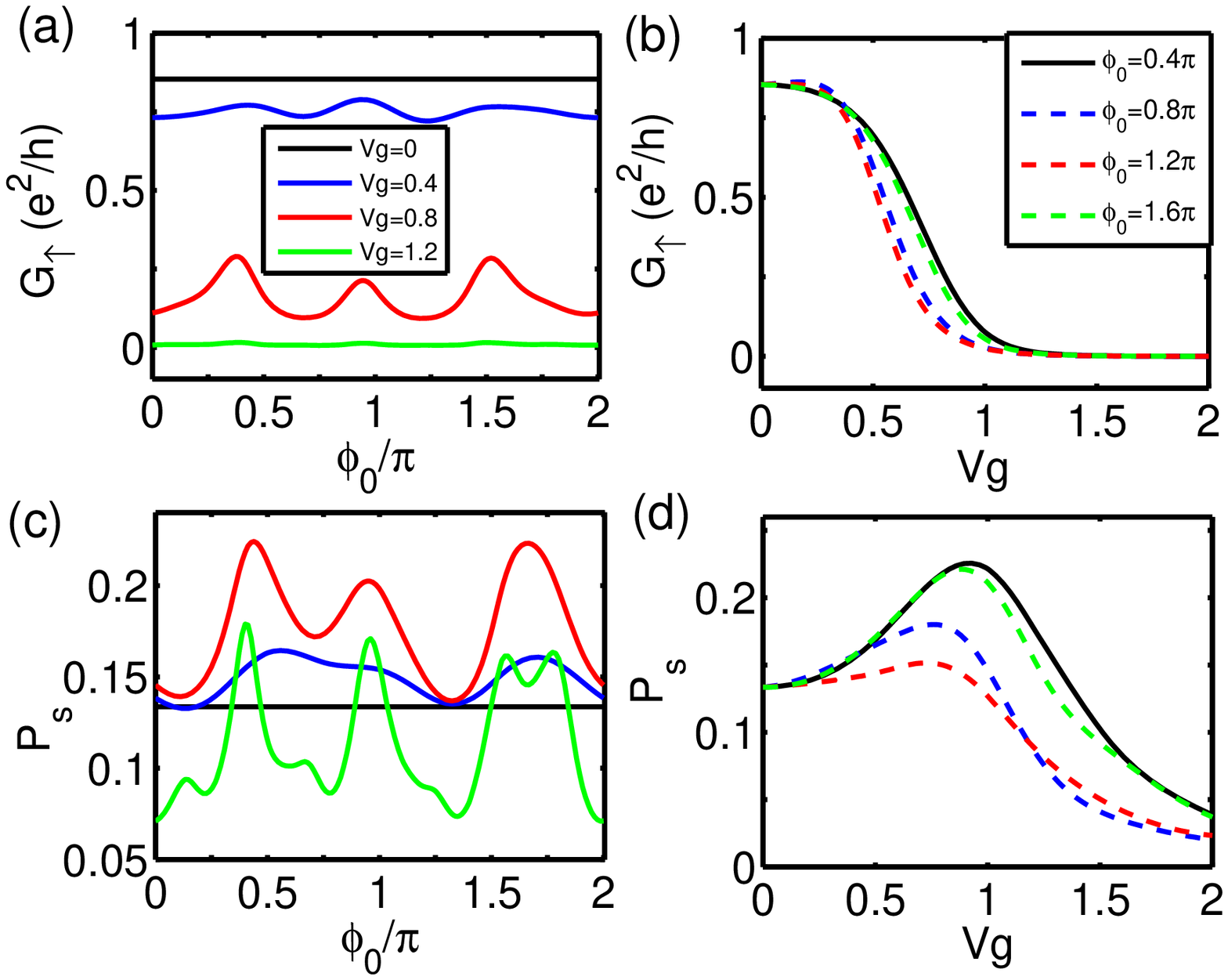}
\includegraphics[scale=0.48]{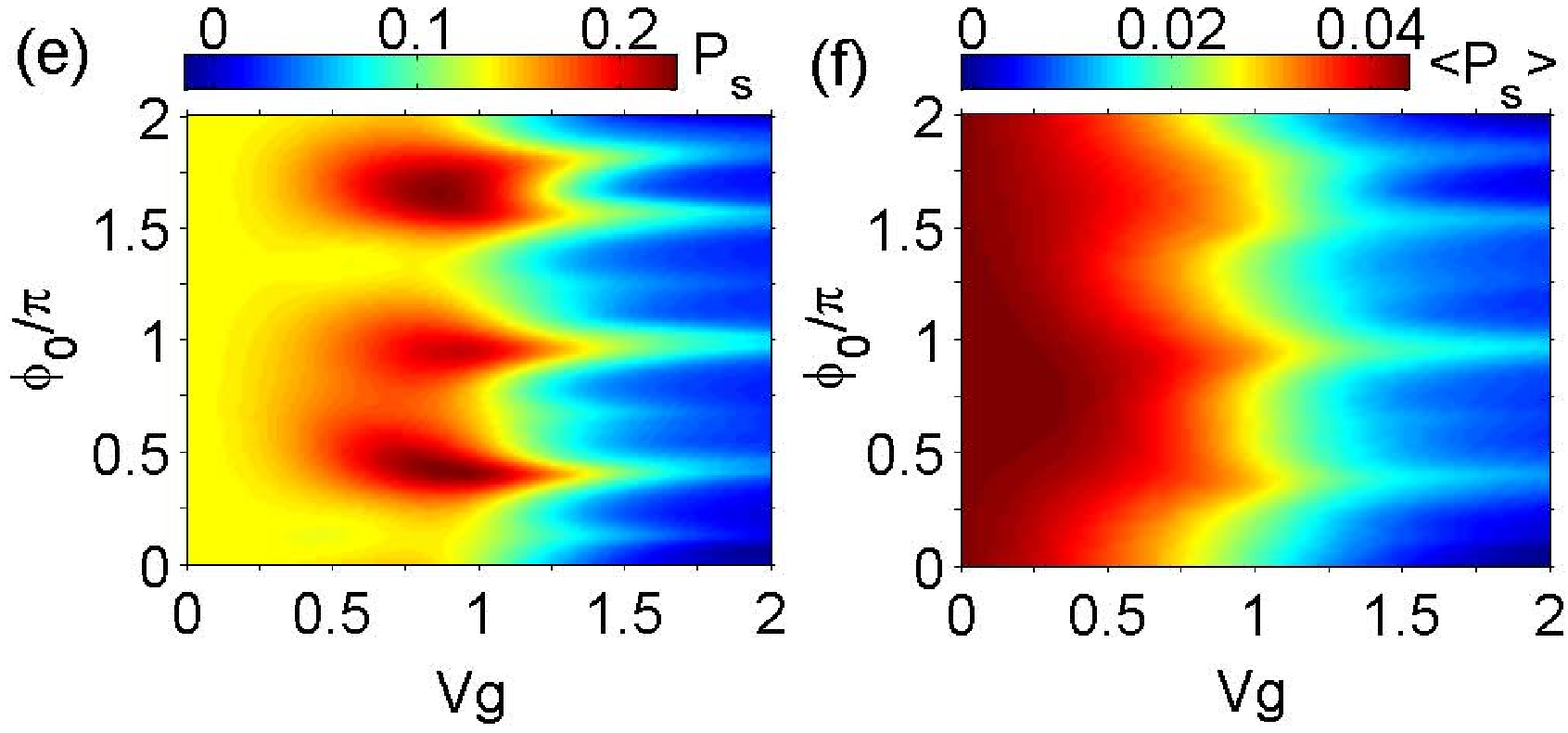}
\caption{\label{fig:two}(color online). (a) $G_\uparrow$ vs $\phi_0$ and (c) $P_s$ vs $\phi_0$ for the peptide with different values of $V_g$. (b) $G_\uparrow$ vs $V_g$ and (d) $P_s$ vs $V_g$ for the peptide with several values of $\phi_0$. Two-dimensional plot of (e) $P_s$ and (f) $\langle P_s\rangle$ vs $V_g$ and $\phi_0$. It is obvious that $P_s$ and $\langle P_s\rangle$ have strong dependence on $V_g$ as well as $\phi_0$. Here, the Fermi energy is $E=-0.41t_1$.}
\end{figure}

Figures~\ref{fig:two}(a) and \ref{fig:two}(c) plot, respectively, the conductance $G_\uparrow$ vs the phase $\phi_0$ and the spin polarization $P_s$ vs the phase $\phi_0$ with four values of the gate voltage $V_g$. It is clear that in the absence of the gate voltage, both $G_\uparrow$ and $P_s$ are independent of $\phi_0$. When the gate voltage is applied, both the curves of $G_\uparrow$-$\phi_0$ and $P_s$-$\phi_0$ present oscillating behaviors. By inspecting Fig.~\ref{fig:two}(a), it can be seen that for relatively small $V_g$ ($V_g=0.4, 0.8$), the curve $G_\uparrow$-$\phi_0$ displays significant oscillation phenomenon with three distinct peaks; while for large $V_g$ ($V_g=1.2$), the curve $G_\uparrow$-$\phi_0$ possesses several peaks with quite small oscillating amplitude, because of the strong gating effect at large $V_g$. As compared with $G_\uparrow$-$\phi_0$, the oscillating amplitude of the curve $P_s$-$\phi_0$ is always very big for all investigated values of $V_g$ and the spin polarization is quite large (Fig.~\ref{fig:two}(c)). One can see that the spin filtration efficiency of the $\alpha$-helical peptide at $V_g=0.4$ and $0.8$ is always larger than that without the gate voltage. Even in the case of large $V_g$, there still exist several intervals of $\phi_0$ at which $P_s$ is larger than that of $V_g=0$. Therefore, the spin filtration efficiency of the $\alpha$-helical peptide could be drastically enhanced by properly adjusting the direction of the gate field.

We then show the conductance $G_\uparrow$ and the spin polarization $P_s$ as a function of the gate voltage $V_g$ with four different values of $\phi_0$, as illustrated in  Figs.~\ref{fig:two}(b) and (d). One can note that the dependence of both $G_\uparrow$ and $P_s$ on $V_g$ is more concise, which is different from the curves $G_\uparrow$-$\phi_0$ and $P_s$-$\phi_0$. The conductance $G_\uparrow$ is almost independent of $V_g$ for $V_g<0.3$ and decreases sharply with increasing $V_g$ for $0.3<V_g<1.1$ (Fig.~\ref{fig:two}(b)). And $G_\uparrow$ tends to zero for $V_g>1.1$, since the transmission valley could appear around the energy $E=-0.41t_1$ due to the strong gating effect. In contrast, the dependence of $P_s$ on $V_g$ is not monotonic and there exists a turning point $V_g^c$ in the curve $P_s$-$V_g$. $P_s$ increases with $V_g$ at first for $V_g<V_g^c$ and is gradually declined by further increasing $V_g$. This phenomenon is irrespective of $\phi_0$, although the turning point may depend upon the direction of the gate field. Furthermore, $P_s$ of the $\alpha$-helical peptide is relatively large for very large gate voltage, e.g., $V_g>1.5$. This indicates that the spin selectivity of the peptide could be enhanced by changing the gate voltage and is robust against the strong gating effect.

\begin{figure}
\includegraphics[scale=0.33]{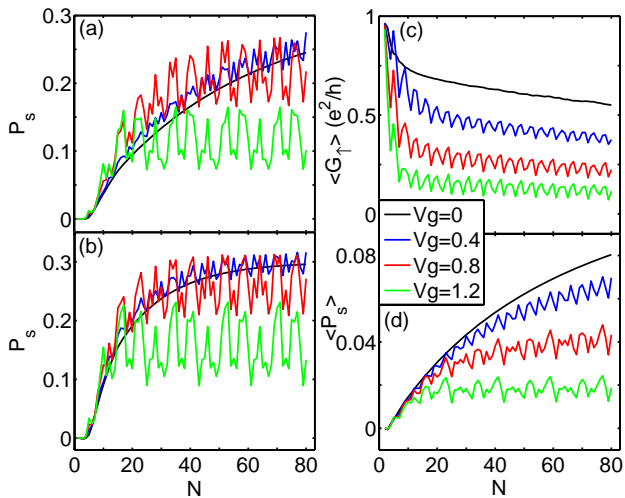}
\caption{\label{fig:four}(color online). (a) $P_s$ vs length $N$ with $\Gamma_d=0.02t_1$ and (b) $P_s$ vs $N$ with $\Gamma_d=0.06t_1$ for the  peptide/protein at $E=-0.41t_1$. (c) Averaged conductance $\langle G_{\uparrow}\rangle$ vs $N$ and (d) $\langle P_s\rangle$ vs $N$ for the  peptide/protein with $\Gamma_d=0.02t_1$. Here, $\phi_0=0.8\pi$, the maximal length $N=80$, and the curves of different colors represent different $V_g$.}
\end{figure}

In the following, we further investigate the spin polarization $P_s$ of the $\alpha$-helical peptide with the magnitude and the direction of the gate field in a wider parameter's range, as illustrated in Fig.~\ref{fig:two}(e). One can see that $P_s$ increases with $V_g$ at first and then is declined by further increasing $V_g$, for almost all values of $\phi_0$. And there exist three extensive domains at which $P_s$ exceeds 20\% and is much bigger than that of $V_g=0$. The approximate range of these three domains is $[0.7, 1.1]$ for $V_g$ and $[0.3\pi, 0.5\pi]$, $[0.9\pi, \pi]$, $[1.5\pi, 1.8\pi]$ for $\phi_0$. Besides, one can identify some other features. (I) For $V_g<0.3$, $P_s$ is nearly independent of $\phi_0$. (II) For $0.3<V_g<1.1$, the dependence of $P_s$ on $\phi_0$ exhibits three peaks. (III) For $V_g>1.1$, the behavior of $P_s$ versus $\phi_0$ becomes complicated and has multiple turning points. The above results reveal that the spin polarization of the peptide is strongly dependent of the magnitude as well as the direction of the gate field, and could be significantly increased by properly tuning the gate voltage. Figure~\ref{fig:two}(f) displays the averaged spin polarization $\langle P_s\rangle$ of the peptide as functions of $V_g$ and $\phi_0$. It can be seen that $\langle P_s\rangle$ monotonically declines with the increase of $V_g$, irrespective of $\phi_0$. $\langle P_s\rangle$ presents oscillating dependence on $\phi_0$ by fixing $V_g$. With the increase of $V_g$, the peak width of $\langle P_s\rangle$-$\phi_0$ decreases and some small peaks appear.

\begin{figure}[b]
\includegraphics[scale=0.4]{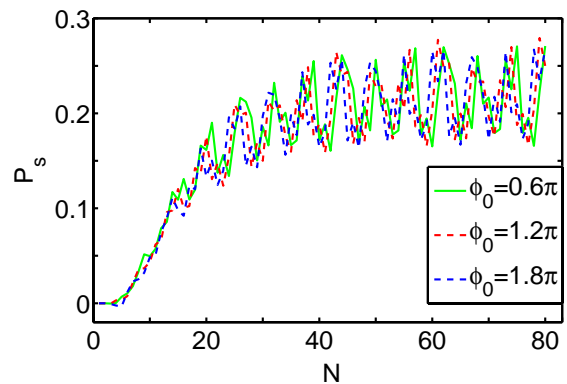}
\caption{\label{fig:five}(color online). Length-dependent $P_s$ for the  peptide/protein with three values of $\phi_0$ at $E=-0.41t_1$ and $V_g=0.8$.}
\end{figure}

Figures~\ref{fig:four}(a) and \ref{fig:four}(b) show the length-dependent spin polarization for four different values of $V_g$ with $\Gamma_d=0.02t_1$ and $\Gamma_d=0.06t_1$, respectively. One can see the following features which are similar for different values of $\Gamma_d$. (1) In the absence of the gate voltage, $P_s$ increases monotonically with $N$.\cite{gam2} For small $\Gamma_d$, the dependence of $P_s$ on $N$ is almost linear and the rising slope decreases very slowly with increasing $N$, which is in excellent agreement with recent experiment.\cite{km} While for relatively large $\Gamma_d$, $P_s$ increases very fast with $N$ for short molecular length and the increasing speed slows down for long molecular length, since larger dephasing strength leads to faster memory loss of the electrons. (2) When the gate voltage is implemented, the curve $P_s$-$N$ displays oscillating behavior due to the gating effect. One could note that the value of the same figurate peak increases quickly with $N$ at first and afterwards is suppressed with the further increase of $N$, which is more obvious for large $V_g$. (3) The oscillating period is length $18$, which corresponds to the same figurate peak value and derives from the fact that the twist angle of the $\alpha$-helical protein is $\Delta\varphi=5\pi/9$. However, the oscillation behavior of the  peptide/protein is much more complicated and several peaks could appear within a period, because a period is composed of five helical circles. This is different from the dsDNA molecule,\cite{gam3} where only one peak exists in a period, since a period is composed of one helical circle in the dsDNA molecule. (4) The oscillating amplitudes of the $P_s$-$N$ curves for $V_g=0.8$ and $V_g=1.2$ are much larger than that for $V_g=0.4$. (5) For all investigated values of the gate voltage, there always exist specific length regions, where $P_s$ under the gate voltage is larger than that without the gate voltage. The range of these specific length regions decreases with increasing $V_g$. For instance, $P_s$ at $V_g=0.4$ is larger than that at $V_g=0$ for almost all the length region. $P_s$ at $V_g=1.2$ is larger than that at $V_g=0$ only for short molecular length. These results further indicate that the spin selectivity of the  peptide/protein could be improved by introducing an external electric field in a quite wide length range, especially for short peptide.

Figures~\ref{fig:four}(c) and \ref{fig:four}(d) plot the averaged spin-up conductance $\langle G_\uparrow \rangle$ versus $N$ and the averaged spin polarization $\langle P_s\rangle$ versus $N$, respectively, with $\Gamma_d=0.02t_1$ for several values of $V_g$. In comparison with the case of $V_g=0$ that both $\langle G_\uparrow \rangle$ and $\langle P_s\rangle$ vary smoothly with $N$, the behaviors of $\langle G_\uparrow \rangle$-$N$ and $\langle P_s\rangle$-$N$ under the gate voltage are oscillating and the oscillating amplitude is almost the same for different values of $V_g$ and $N$. It can be seen that the peak value of $\langle G_\uparrow \rangle$-$N$ declines fast with $N$ at first and is suppressed for long molecular length, especially in the case of large gate voltage. Contrary to the dependence of $\langle G_\uparrow \rangle$ on $N$, $\langle P_s\rangle$ increases monotonically with $N$ in the absence of gate voltage, which is consistent with the experiment.\cite{km} While for $V_g \neq 0$ in the $\langle P_s\rangle$-$N$ curve, the value of the same figure peak increases with $N$ at first and then tends to saturation by further increasing $N$, especially in the case of large $V_g$. The larger $V_g$ corresponds to the smaller values of $\langle G_\uparrow \rangle$ and $\langle P_s\rangle$, which is independent of $N$, because of the gating effect. What's more, both the averaged conductance $\langle G_\uparrow \rangle$ and the averaged spin polarization $\langle P_s\rangle$ are still comparatively large for $N=80$ and $V_g=1.2$.

Figure~\ref{fig:five} displays the length-dependent spin filtration efficiency of the $\alpha$-helical peptide/protein for different values of the phase $\phi_0$ at $\Gamma_d=0.02t_1$. $P_s$ oscillates by increasing $N$ in the presence of the gate voltage, for whatever the value of $\phi_0$. The curve $P_s$-$N$ for various $\phi_0$ presents two similar oscillating features as that of the $P_s$-$N$ curve for different $V_g$. Firstly, the same figurate peak shows that the period of the length is $18$ amino acids. Secondly, the value of the same figurate peak increases with $N$ at first and then is suppressed by further increasing $N$. Although the oscillating amplitude of the curve $P_s$-$N$ nearly shows independence on $\phi_0$, the position of the same figurate peak moves towards larger $N$ with $\phi_0$ at first and then is shifted towards smaller $N$ by the further increase of $\phi_0$. For instance, one local maximum of the corresponding peak locates at $N=68$, $74$, and $73$, by increasing $\phi_0$ from $0.6\pi$, $1.2\pi$, to $1.8\pi$. It can also be seen that there exists a quite wide length range where $P_s$ varies drastically with $\phi_0$ by fixing $N$. The results imply that the spin polarization of the peptide/protein for various length is $\phi_0$-dependent and could be tuned by  adjusting the direction of the gate field.

\begin{figure}
\includegraphics[scale=0.4]{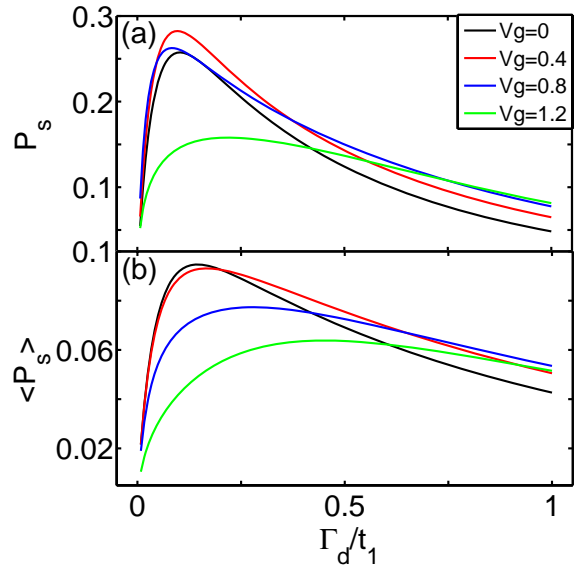}
\caption{\label{fig:six}(color online). (a) $P_s$ vs dephasing strength $\Gamma_d$ at $E=-0.41t_1$ and (b) $\langle P_s \rangle $ vs $\Gamma_d$ for the peptide with various values of $V_g$ by fixing $\phi_0=0.8\pi$.}
\end{figure}

The dephasing occurs inevitably in the experiment and its strength may rely on various experimental conditions, such as the temperature. Here, we consider the influence of the dephasing on the spin transport through the $\alpha$-helical peptide under the gate voltage. Figures~\ref{fig:six}(a) and \ref{fig:six}(b) plot the spin polarization $P_s$ and $\langle P_s\rangle$, respectively, as a function of $\Gamma_d$ with the strength up to $\Gamma_d=t_1$ for four values of $V_g$. It is clear that there exists a crossover of $\Gamma_d$ in each curve of $P_s$-$\Gamma_d$ and $\langle P_s\rangle$-$\Gamma_d$, although the specific value of this crossover may depend on the gate voltage. Both $P_s$ and $\langle P_s\rangle$ increase quickly with $\Gamma_d$ in the weak dephasing regime and then are slowly declined by further increasing $\Gamma_d$ in the strong dephasing regime. This phenomenon does not depend upon the value of the gate voltage and originates from the two competitive effects of the dephasing.\cite{gam2} On the one hand, the dephasing promotes the openness of the system because of the coupling to the B\"{u}ttiker's virtual lead and produces the spin asymmetry. On the other hand, the dephasing gives rise to the memory loss of the electrons and shrinks the spin polarization. When the former effect dominates, the spin polarization could be enhanced by increasing $\Gamma_d$; when the latter effect prevails, the spin polarization will be declined. Besides, one can see from the curve $\langle P_s\rangle$-$\Gamma_d$ that the position of the crossover is shifted toward larger $\Gamma_d$ by increasing $V_g$ and $\langle P_s\rangle$ at the crossover is declined. However, in the strong dephasing regime where the spin polarization decreases with increasing $\Gamma_d$, the absolute value of the decreasing slope is declined by increasing $V_g$. This indicates that the spin filtration efficiency could be more robust against the dephasing in the region of large $V_g$. In other words, both $P_s$ and $\langle P_s\rangle$ could be enhanced by the gate voltage in the strong dephasing regime. Therefore, the gate voltage could enhance the spin filtration efficiency of the $\alpha$-helical peptide in a wide dephasing range, especially in the case of the strong dephasing.

\begin{figure}
\includegraphics[scale=0.285]{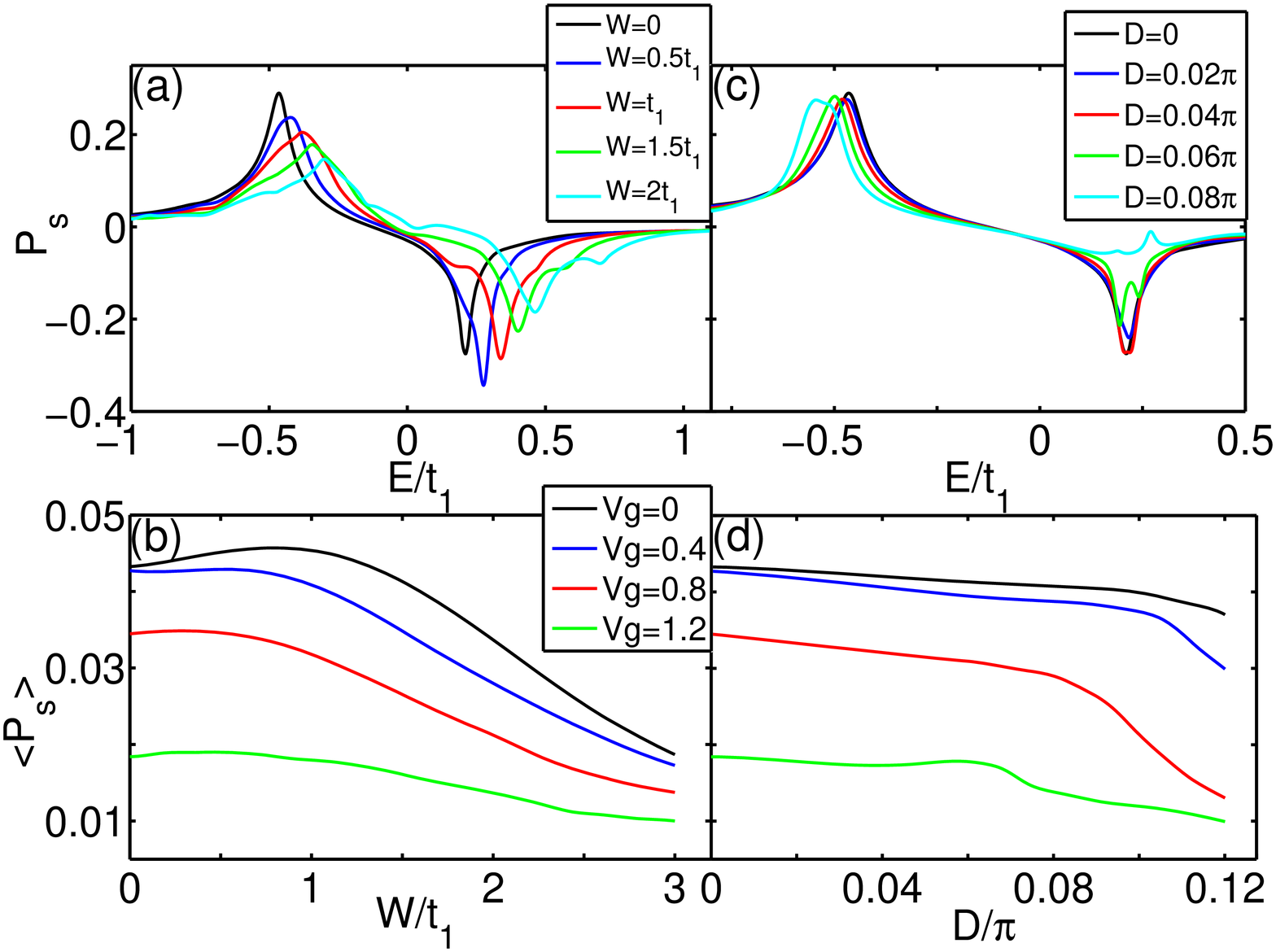}
\caption{\label{fig:seven}(color online). (a) $P_s$ vs $E$ under the on-site energy disorder with several strength $W$ and (c) $P_s$ vs $E$ under the space angle disorder with several strength $D$ for the peptide by fixing $V_g =0.4$ and $\phi_0=0.8\pi$. (b) $\langle P_s\rangle$ vs on-site energy disorder strength $W$ and (d) $\langle P_s\rangle$ vs space angle disorder strength $D$ for the peptide with different $V_g$ at $\phi_0=0.8\pi$. All the data are performed for a single disorder configuration and similar results can be also obtained with other disorder configurations.}
\end{figure}

Distinct amino acids that form the $\alpha$-helical peptide/protein could lead to the on-site energy disorder. Next, we study the influence of the on-site energy disorder on the spin transport through the peptide molecule in the presence of the gate voltage. A stochastic variable $w_n$ is added to each $\varepsilon_ {n}$ to simulate random on-site energy disorder with $w_n$ uniformly distributed in the region $[-W/2, W/2]$ and $W$ is the disorder strength. Figure~\ref{fig:seven}(a) plots the spin polarization $P_s$ versus the energy $E$ for five different values of $W$ with $V_g=0.4$ and $\phi_0 = 0.8\pi$. One notes that the positions of both the peak and the valley move towards higher energy with increasing $W$ and the corresponding $P_s$ is usually declined, due to the disorder-induced Anderson localization effect. However, $P_s$ at some energy regions could be enhanced by increasing $W$ and is robust against the on-site energy disorder in the presence of the gate voltage. This statement is further demonstrated in Fig.~\ref{fig:seven}(b), where  $\langle P_s\rangle$ is displayed as a function of $W$ for several values of $V_g$. It clearly appears that $\langle P_s \rangle$ could be increased by increasing $W$ in the weak disorder region and is still considerably large in the strong disorder region.

Each amino acid will fluctuate around its equilibrium position, which can be simulated by introducing a random variable $d_n$ to each $\theta_{n1}$ between the $n$th and $(n+1)$th sites. $d_n$ is evenly distributed within the range $[-D/2, D/2]$ with $D$ the disorder strength of the space angle. Herein, we assume that all of the amino acids locate at the side surface of the cylinder whose radius $R$ is constant and the Euclidean distance $l_{n1}$ is always the same as $l_1$.\cite{gam3} Then, the stacking distance and the twist angle between the $n$th and $(n+1)$th amino acids is written as $h_{n+1}-h_n= l_1\sin(\theta_{n1})$ and $\varphi_{n+1}- \varphi_{n} =2\arcsin[l_1 \cos (\theta_{n1})/(2R)]$. Figures~\ref{fig:seven}(c) and \ref{fig:seven}(d) investigate the effect of the space angle disorder on the spin polarization $P_s$ and $\langle P_s\rangle$ of the peptide in the presence of the gate voltage. It can be seen that the peak of $P_s$-$E$ is shifted towards the lower energy with the increase of $D$ and the peak value nearly remains unchanged. Besides, $\langle P_s\rangle$ decreases quite slowly with increasing $D$ within a wide range of the space angle disorder and remains large in the case of large $D$ and $V_g$. From the above, we can conclude that the spin filtration efficiency of the $\alpha$-helical peptide is robust against both the on-site energy disorder and the space angle disorder under the gate voltage.

\section{\label{sec4}Conclusions}

In summary, we investigate the influence of the gate voltage on the quantum spin transport along the $\alpha$-helical peptide/protein molecule contacted by two nonmagnetic electrodes. The spin filtration efficiency of the $\alpha$-helical peptide/protein can be enhanced significantly by modulating the magnitude as well as the direction of the gate field, and is robust against the dephasing, the on-site  energy disorder and the space angle disorder in the presence of the gate voltage. A constructive scheme is provided for further experimental studies on protein spintronics and meanwhile, could be readily carried out and checked.

\section*{Acknowledgments}

This work was supported by NBRP of China (2012CB921303 and 2015CB921102), NSF-China under Grants No. 11274364,
and the Fundamental Research Funds for the Central Universities under Grant No. AUGA5710013615.

\end{document}